\documentclass[12pt]{article}
\usepackage{amssymb,amsmath,epsfig}

\begin{document}

\title{\bf Spherical Thin-Shell Wormholes and Modified Chaplygin Gas}

\author{M. Sharif$^1$ \thanks{msharif.math@pu.edu.pk} and M. Azam$^{1,2}$
\thanks{azammath@gmail.com}\\
$^1$ Department of Mathematics, University of the Punjab,\\
Quaid-e-Azam Campus, Lahore-54590, Pakistan.\\
$^2$ Division of Science and Technology, University of Education,\\
Township Campus, Lahore-54590, Pakistan.}

\date{}

\maketitle
\begin{abstract}
The purpose of this paper is to construct spherical thin-shell
wormhole solutions through cut and paste technique and investigate
the stability of these solutions in the vicinity of modified
Chaplygin gas. The Darmois-Israel formalism is used to formulate the
stresses of the surface concentrating the exotic matter. We explore
the stability of the wormhole solutions by using the standard
potential method. We conclude that there exist more stable as well
as unstable solutions than the previous study with generalized
Chaplygin gas \cite{15}.
\end{abstract}
{\bf Keywords:} Thin-shell wormholes; Darmois-Israel conditions; Stability.\\
{\bf PACS:} 04.20.Gz; 04.20.-q; 04.40.Nr; 04.70.Bw.

\section{Introduction}

Wormhole physics has been an interesting subject for physicists
since the first traversable wormhole by Morris and Throne as a
solution of the Einstein field equations \cite{1}. The fascinating
idea was to connect two distinct or the same universe through a
handel or tunnel \cite{1,2}. The main issue with this wormhole is
the existence of inevitable amount of exotic matter around the
throat. The problem of minimizing the usage of exotic matter for the
physically viability of wormhole has received a considerable
attention. For instance, it was shown that with suitable choice of
wormhole geometry \cite{3}, it is possible to reduce exotic matter
around the wormhole throat. Visser \cite{2,4} analyzed that the
violation of energy condition could be minimized with the
construction of thin-shell wormholes through cut and paste
technique. This construction restricts exotic matter to be placed at
the wormhole throat. The study of thin-shell wormholes with the
Darmois-Israel formalism \cite{5,6} has widely been discussed in
literature \cite{6a}-\cite{6d}.

It is believed that any traversable wormhole is of physical interest
if it is stable under linear perturbations preserving the
spherically symmetry. In this scenario, many authors carried out the
stability analysis of thin-shell wormholes through linear
perturbations. Poisson and Visser \cite{7} performed stability
analysis for the Schwarzschild thin-shell wormholes. Afterwards, the
same analysis was extended for charge \cite{8} and cosmological
constant \cite{9}. It was found that the stability regions would be
increased with the inclusion of charge and positive cosmological
constant while decreased for the negative cosmological constant. The
dilaton thin-shell wormholes with and without charge was
investigated by different authors \cite{10}. Eiroa and his
collaborators \cite{11} constructed cylindrical thin-shell wormholes
and investigated their stability. Recently, we have explored the
stability of spherical and cylindrical thin-shell wormholes under
linear perturbations \cite{12}.

The choice of equation of state for the description of matter
present in the wormhole throat has a great relevance in the
existence and stability of wormhole static solutions. In this
context, Eiroa and Simeone \cite{13} constructed the spherical
thin-shell wormholes with matter source as pure Chaplygin gas. They
concluded that static stable and unstable wormhole solutions exist
depending upon the parameters in the model. Bandyopadhyay et al.
\cite{14} generalized this analysis by considering simple modified
Chaplygin gas and found that stable static wormhole solutions are
also possible in the Schwarzschild as well as Schwarzschild
de-Sitter cases. Later, Eiroa \cite{15} formulated thin-shell
wormholes in the scenario of generalized Chaplygin gas and found
more unstable static solutions. Also, Gorini et al. \cite{15a,15b}
investigated the Tolman-Oppenheimer-Volkoff equations and found the
wormhole like solutions with a spacetime singularity for a model
filled with the Chaplygin gas and generalized Chaplygin gas (GCG).

It has been established that our universe is dominated by dark
energy which is responsible for the continuous acceleration of the
universe. There are many candidates of dark energy such as Chaplygin
gas, GCG and modified Chaplygin gas (MCG), etc.  In this scenario,
Kamenshchik et al. \cite{15c} discussed the first ever FRW
cosmological model supported by the Chaplygin gas and GCG. The same
others \cite{15d} explored branes in the background of BTZ and anti
de-Sitter Schwarzschild black holes in the context of MCG. The
equation of state for the MCG is defined as \cite{16}
\begin{equation}\label{a}
p=A{\sigma}-\frac{B}{\sigma^\beta},
\end{equation}
where $A>0,~B>0$ and $0<{\beta}\leq1$. A class of Chaplygin gas can
be recovered for different choices of parameters $A,~B$ and $\beta$,
such as:
\begin{itemize}
\item{for $A=0,~\beta=1$, we get usual Chaplygin gas.}
\item{for $A=0$, it reduces to GCG.}
\item{for $\beta=1$, it is another form of simple MCG.}
\end{itemize}
In this paper, we construct spherical thin-shell wormholes with MCG
as matter located on the shell. The paper is organized as follows.
The next section is devoted to provide a general formalism for the
construction of spherical thin-shell wormholes. Section \textbf{3}
deals with the procedure about the stability of static wormhole
solutions which is applied to particular examples in section
\textbf{4}. Finally, we conclude the results in the last section.

\section{Basic Formalism for Thin-Shell Wormhole}

In this section, we develop a general formalism to construct
spherical thin-shell wormholes. We take general spherically
symmetric spacetime given by
\begin{equation}\label{1}
ds^2=-N(r)dt^{2}+N^{-1}(r)dr^{2}+G(r)(d\theta^{2}
+\sin^2\theta{d\phi^2}),
\end{equation}
where $0\leq{\theta}\leq{\pi},~ 0\leq{\phi}<{2\pi}$ are the angular
coordinates and $N(r),~G(r)$ are positive functions of the radial
coordinate $r>0$. The cut and paste technique is an elegant way to
construct a thin-shell wormhole solution. For this purpose, we cut
the interior region of the manifold (\ref{1}) with $r<a$, yielding
two identical four-dimensional copies $\mathcal{V}^{\pm}$ with
radius $r\geq{a}$
\begin{equation}\label{3}
\mathcal{V}^{\pm}=\{x^{\mu}=(t,r,\theta,\phi)/r\geq{a}\}.
\end{equation}
The assumed radius $``a"$ is taken greater than the horizon radius
$r_h$ of the manifold (\ref{1}) to avoid singularities and horizons.
We obtain a new manifold by pasting these copies at the timelike
hypersurface $\Sigma=\Sigma^\pm=\{r-a=0\}$, where the boundaries
$\Sigma^\pm$ correspond to $\mathcal{V}^{\pm}$. This created
manifold $\mathcal{V}=\mathcal{V}^{+}\cup{\mathcal{V}^{-}}$ is
called geodesically complete satisfying the flare-out condition
i.e., $G'(a)>0$, which describes a wormhole having two regions stick
with a throat radius $a$ (called minimal surface area). The proper
radial distance can be defined on $\mathcal{V}$ as
$s=\pm\int^r_a{\sqrt{\frac{1}{N(r)}}}dr$, which depicts the throat
position for $s=0$, where $\pm$ correspond to ${\mathcal{V}^\pm}.$

The intrinsic three-dimensional metric at the throat $\Sigma$ is
given by
\begin{equation}\label{4}
ds^2=-d\tau^2+a^2(\tau)(d\theta^2+\sin^2\theta{d\phi^2}),
\end{equation}
where $a$ is a function of proper time $\tau$. We have applied the
Darmois-Israel formalism to the matter at $\Sigma$. The extrinsic
curvature $K^{\pm}_{ij}$ associated with the shell is defined as
\begin{equation}\label{5}
K^{\pm}_{ij}=-n^{\pm}_{\gamma}\left(\frac{{\partial}^2x^{\gamma}_{\pm}}
{{\partial}{\eta}^i{\partial}{\eta}^j}+{\Gamma}^{\gamma}_{{\mu}{\nu}}
\frac{{{\partial}x^{\mu}_{\pm}}{{\partial}x^{\nu}_{\pm}}}
{{\partial}{\eta}^i{\partial}{\eta}^j}\right),\quad(i,~j=0,2,3).
\end{equation}
where $4$-vector unit normals $n^{\pm}_{\gamma}$ to
$\mathcal{V}^{\pm}$ are
\begin{equation}\label{6}
n^{\pm}_{\gamma}=\pm\left|g^{\mu\nu}\frac{\partial{f}}{\partial{x^{\mu}}}
\frac{\partial{f}}{\partial{x^{\nu}}}\right|^{-\frac{1}{2}}\frac{\partial{f}}{\partial{x^\gamma}}
=\left(-\dot{a},\frac{\sqrt{N(r)+\dot{a}^2}}{N(r)},0,0\right),
\end{equation}
satisfying the relation $n^{\gamma}n_{\gamma}=1$. Using the
orthonormal basis for Eq.(\ref{1}),
$\{e_{\hat{\tau}}=e_{\tau},~e_{\hat{\theta}}=[G(a)]^{-\frac{1}{2}}e_{\theta},~
e_{\hat{\phi}}=[G(a)\sin^2{\theta}]^{-\frac{1}{2}}e_{\phi}\}$, we
obtain the following non-trivial extrinsic curvature components
\begin{equation}\label{7}
K^{\pm}_{\hat\tau\hat\tau}=\mp\frac{N'(a)+2\ddot{a}}{2\sqrt{N(a)+\dot{a}^2}},
\quad
K^{\pm}_{\hat\theta\hat\theta}=K^{\pm}_{\hat\phi\hat\phi}={\pm}\frac{G'(a)}{2G(a)}\sqrt{N(a)+\dot{a}^2}.
\end{equation}
where dot and prime mean derivatives with respect to $\tau$ and $r$,
respectively. The discontinuity in the extrinsic curvatures across a
junction surface yields the Lanczos equations on the shell
\begin{equation}\label{8}
S_{\hat{i}\hat{j}}=\frac{1}{8\pi}\left\{g_{\hat{i}\hat{j}}K-[K_{\hat{i}\hat{j}}]\right\},
\end{equation}
where $S_{\hat{i}\hat{j}}=diag(\sigma,p_{\hat\theta},p_{\hat\phi})$
is the surface stress-energy tensor which determines the surface
stresses of $\Sigma$, and
$[K_{\hat{i}\hat{j}}]=K^{+}_{\hat{i}\hat{j}}-K^{-}_{\hat{i}\hat{j}}$
$K=tr[K_{\hat{i}\hat{j}}]=[K^{\hat{i}}_{\hat{i}}]$ provides
relations between the extrinsic curvatures. The surface stresses
i.e., surface energy density $\sigma$ and surface pressures
$p=p_{\hat\theta}=p_{\hat\phi}$ to shell with Eqs.(\ref{7}) and
(\ref{8}) become
\begin{eqnarray}\label{9}
\sigma&=&-\frac{G'(a)}{4\pi{G(a)}}\sqrt{N(a)+\dot{a}^2},\\\label{10}
p&=&p_{\hat\theta}=p_{\hat\phi}=\frac{\sqrt{N(a)+\dot{a}^2}}{8\pi}
\left[\frac{2\ddot{a}+N'(a)}{N(a)+\dot{a}^2}+\frac{G'(a)}{G(a)}\right].
\end{eqnarray}
Inserting the above equations in Eq.(\ref{a}), we obtain a second
order differential equation describing the evolution of the wormhole
throat
\begin{eqnarray}\nonumber
&&\left\{\left[2\ddot{a}+N'(a)\right]G(a)+\left[\left(N(a)+\dot{a}^2\right)
\left(1+2A\right)\right]G'(a)\right\}\left[G'(a)\right]^{\beta}\\\label{11}&-&2B(4\pi{G(a)})^{1+\beta}
\left[N(a)+\dot{a}^2\right]^\frac{1-\beta}{2}=0.
\end{eqnarray}

\section{Stability Analysis and Linear Perturbations}

In this section, we investigate stability of wormhole static
solutions through linear perturbations \cite{10}. We consider static
configuration $(\dot{a}=\ddot{a}=0)$ of surface stresses (energy
density, surface pressure) and  dynamical equation of wormhole from
Eqs.(\ref{9})-(\ref{11}) as
\begin{eqnarray}\label{12}
\sigma_0=-\frac{\sqrt{N(a_0)}}{4\pi}\frac{G'(a_0)}{G(a_0)},\quad
p_0=\frac{\sqrt{N(a_0)}}{8\pi}\left[\frac{N'(a_0)}{N(a_0)}+\frac{G'(a_0)}{G(a_0)}\right],
\end{eqnarray}
\begin{eqnarray}\nonumber
&&\left\{G(a_0)N'(a_0)+G'(a_0)N(a_0)
\left(1+2A\right)\right\}\left[G'(a_0)\right]^{\beta}\\\label{13}&-&2B(4\pi{G(a_0)})^{1+\beta}
\left[N(a_0)\right]^\frac{1-\beta}{2}=0.
\end{eqnarray}
The conservation equation with Eqs.(\ref{9}) and (\ref{10}) can be
defined as
\begin{eqnarray}\label{14}
\frac{d}{d\tau}(\sigma{\Delta})+p\frac{d\Delta}{d\tau}=
\left\{[G'(a)]^2-2G(a)G''(a)\right\}\left\{\frac{\dot{a}\sqrt{N(a)+\dot{a}^2}}{2G(a)}\right\},
\end{eqnarray}
where $\Delta=4\pi{G(a)}$ gives the wormhole throat area. The left
hand side of the above equation describes the rate of change of
throat internal energy and throat's internal forces work done. Using
wormhole throat area definition and Eq.(\ref{9}), the above equation
becomes
\begin{eqnarray}\label{15}
G(a)\dot{\sigma}+G'(a)\dot{a}(\sigma+p)=-\left\{[G'(a)]^2-2G(a)G''(a)\right\}
\left\{\frac{\dot{a}\sigma}{2G'(a)}\right\},
\end{eqnarray}
which further reduces to
\begin{equation}\label{16}
G(a){\sigma}'+G'(a)(\sigma+p)+\left\{[G'(a)]^2-2G(a)G''(a)\right\}
\left\{\frac{\sigma}{2G'(a)}\right\}=0,
\end{equation}
where we have used ${\sigma}'=\frac{\dot{\sigma}}{\dot{a}}$. It is
noted from Eq.(\ref{a}) that $p$ is a function of $\sigma$. Thus,
the above equation is a first order differential equation in
$\sigma(a)$ and can be written as $\sigma'(a)=H(a,\sigma(a))$. This
equation yields a unique solution for a given initial condition such
that $H$ has continuous partial derivative. The integration of
Eq.(\ref{16}) gives $\sigma(a)$, thus the thin-shell equation of
motion describing the throat dynamics can be obtained from
Eq.(\ref{9}) as
\begin{equation}\label{17}
\dot{a}^2+\Phi(a)=0.
\end{equation}
where the potential function $\Phi(a)$ is defined by
\begin{equation}\label{18}
\Phi(a)=N(a)-16\pi^2\left[\frac{G(a)}{G'(a)}{\sigma(a)}\right]^2.
\end{equation}
For the stability analysis of static solutions under the radial
perturbations, we expand the potential function $\Phi(a)$ by Taylor
series expansion around $a=a_0$ upto second order as
\begin{eqnarray}\label{19}
\Phi(a)=\Phi(a_0)+\Phi'(a_0)(a-a_0)+\frac{1}{2}\Phi''(a_0)(a-a_0)^2+O[(a-a_0)^3].
\end{eqnarray}
The existence and stability of static solution depend upon the
inequalities $a_0>r_h$ and $\Phi''(a_0)\lessgtr{0},~
\Phi(a_0)=0=\Phi'(a_0)$, respectively. The solution will be stable
or unstable, if $\Phi''(a_0)>0$ or $\Phi''(a_0)<0$, respectively.
The first derivative of potential function with Eq.(\ref{16})
becomes
\begin{equation}\label{21}
\Phi'(a)=N'(a)+16{\pi}^2\sigma(a)\frac{G(a)}{G'(a)}
\left[\sigma(a)+2p(a)\right].
\end{equation}
Also, from Eq.(\ref{a}), we have
\begin{equation}\label{21a}
p'(a)={\sigma'(a)}\left[(1+\beta)A-\frac{\beta{p(a)}}{\sigma(a)}\right],
\end{equation}
and would be written as
\begin{equation}\label{23}
{\sigma'(a)}+2p'(a)={\sigma}'(a)\left[1+2\{(1+\beta)A-\frac{\beta{p(a)}}{\sigma(a)}\}\right].
\end{equation}
Using the above equation and Eq.(\ref{16}), the second derivative of
potential function yields
\begin{eqnarray}\nonumber
\Phi''(a)&=&N''(a)-8{\pi}^2\left\{[\sigma(a)+2p(a)]^2+2\sigma(a)
\left[\left(\frac{3}{2}-\frac{G(a)G''(a)}{[G'(a)]^2}\right)\sigma(a)
\right.\right.\\\label{24}&+&\left.\left.p(a)\right]
\left[1+2\left((1+\beta)A
-\frac{\beta{p(a)}}{\sigma(a)}\right)\right]\right\}.
\end{eqnarray}
Using Eq.(\ref{12}), it follows that
\begin{eqnarray}\nonumber
\Phi''(a_0)&=&N''(a_0)+\frac{(\beta-1)[N'(a_0)]^2}{2N(a_0)}+{N'(a_0)}
\left[(1-\beta+2A(1+\beta))\right.\\\nonumber&\times&\left.\frac{G'(a_0)}{2G(a_0)}+
\beta\frac{G''(a_0)}{G'(a_0)}\right]
+(1+2A)(1+\beta)\left[\frac{G''(a_0)}{G(a_0)}\right.\\\label{25}&-&\left.
\left(\frac{G'(a_0)}{G(a_0)}\right)^2\right]N(a_0).
\end{eqnarray}

\section{Applications to Particular Examples}

In this section, we apply the general formalism developed in the
above sections for the construction and stability of spherical
thin-shell wormholes to Reissner-Nordstr\"{o}m (RN), Schwarzschild
de-Sitter and anti de-Sitter geometries.

\subsection{Reissner-Nordstr\"{o}m Wormholes}

Here, we formulate the RN wormholes and investigate their stability.
The metric functions for RN have the form
\begin{eqnarray}\label{26}
N(r)=1-\frac{2M}{r}+\frac{Q^2}{r^2},\quad G(r)=r^2,
\end{eqnarray}
where $M,~Q$ are the mass and charge associated with the RN
geometry. The inner $(-) $ and outer $(+)$ horizons for $0<|Q|<M$
turn out to be
\begin{eqnarray}\label{27}
r^{\pm}=M\pm{\sqrt{M^2-Q^2}}.
\end{eqnarray}
The extremal black hole is obtained for $|Q|=M$, while a naked
singularity is formed for $|Q|>M$. For the RN wormhole, the
condition $a_0>r_h=r^{+}$ must be satisfied. The surface energy
density and surface pressure for the RN wormhole solution becomes
\begin{eqnarray}\label{28}
\sigma_0=-\frac{\sqrt{a^2_0-2Ma_0+Q^2}}{2\pi{a^2_0}},\quad
p_0=\frac{a_0-M}{4\pi{a_0}\sqrt{a^2_0-2Ma_0+Q^2}}.
\end{eqnarray}
The dynamical equation satisfied by the throat radius is obtained by
substituting Eq.(\ref{26}) in (\ref{13})
\begin{eqnarray}\nonumber
&&a^2_0(1+2A)-(1+4A)a_0M+2AQ^2\\\label{29}
&&-2B(2\pi)^{1+\beta}a^{2+2\beta}_0
(a^2_0-2Ma_0+Q^2)^{\frac{1-\beta}{2}}=0.
\end{eqnarray}
Also, the potential function with Eqs.(\ref{25}) and (\ref{26}) can
be written as
\begin{eqnarray}\nonumber
\Phi''(a_0)&=&\frac{1}{a^4_0(a^2_0-2Ma_0+Q^2)}\left\{-2\left
[(1+\beta)\left(a^4_0(1+2A)+4AQ^4\right.\right.\right.\\\nonumber
&-&\left.\left.\left.10Aa^3_0M-14a_0AMQ^2+(3+12A)a^2_0M^2+6Aa^2_0Q^2\right)
\right.\right.\\\label{30}
&-&\left.\left.a^3M(3+4\beta)-a_0MQ^2(1+2\beta)+2\beta{a^2_0}Q^2\right]\right\}.
\end{eqnarray}
\begin{figure}
\centering \epsfig{file=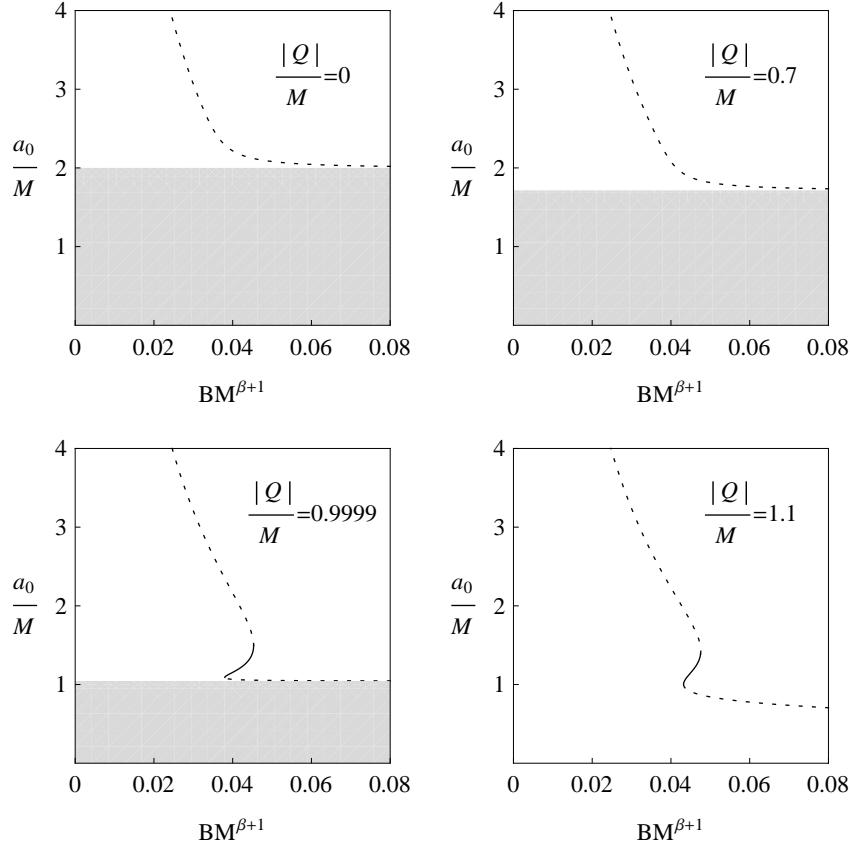}\caption{RN thin-shell wormholes
under radial perturbations with $\beta=0.2,~A=1$ and different
values of $\frac{|Q|}{M}$. The solid and dotted curves represent
stable and unstable solutions respectively, and the grey zones
correspond to the non-physical regions, where $a_0\leq{r_h}$.}
\end{figure}
We solve Eq.(\ref{29}) for $a_0$ numerically with different values
of $0<\beta\leq{1}$. The resulting solution is substituted in
Eq.(\ref{30}) in order to check whether the solution is stable or
unstable depending upon $\Phi''(a_0)>0$ or $\Phi''(a_0)<0$,
respectively. The stable and unstable static solutions are
represented by the black and dotted curves, respectively. We can
summarize the results in Figures \textbf{1-3} for RN wormholes as
follows:
\begin{itemize}
\item{In Figure \textbf{1} when $\beta=0.2$, there exists one unstable static solution
corresponding to $\frac{|Q|}{M}=0,~0.7$ and two unstable and one
stable solution for $\frac{|Q|}{M}=0.999$. The throat radius in each
case decreases for large value of $BM^{\beta+1}$ which touches the
horizon radius of the given manifold. Also, the horizon radius
decreases and eventually disappears for large values of charge. For
$\frac{|Q|}{M}=1.1,$ there are three solutions two of them are
unstable and one is stable.}
\item{When $\beta=0.6$ (Figure \textbf{2}), there exists one unstable static solution
corresponding to $\frac{|Q|}{M}=0,~0.7,~1.1$, while two unstable and
one stable solution for $\frac{|Q|}{M}=0.999$. The throat radius has
similar behavior as in the above case.}
\item{When $\beta=1$ (Figure \textbf{3}), there exist both stable
and unstable static solutions for $0<\frac{|Q|}{M}<1$, while only
unstable solution for $\frac{|Q|}{M}>1$.}
\end{itemize}

We would like to mention here that our results for $\beta=0.2$ are
similar to that reported in \cite{15} for
$\frac{|Q|}{M}=0,~0.7,~0.999$, whereas for $\frac{|Q|}{M}=1.1$, we
have one extra unstable solution. When $\beta=0.6$, we have only one
unstable solution for $\frac{|Q|}{M}=1.1$. Similarly, for $\beta=1$,
we have one more stable solution for $\frac{|Q|}{M}=0,~0.7$ and only
unstable for $\frac{|Q|}{M}=1.1$.
\begin{figure}
\centering \epsfig{file=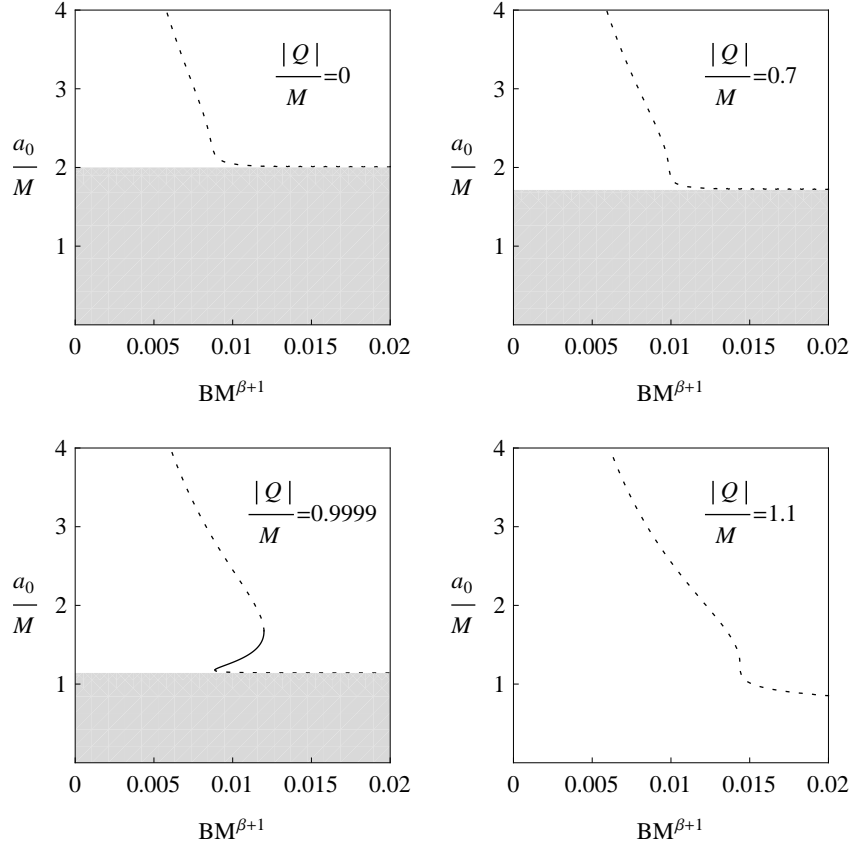}\caption{RN thin-shell wormholes for
$\beta=0.6,~A=1$ with different values of $\frac{|Q|}{M}$.}
\end{figure}
\begin{figure}
\centering \epsfig{file=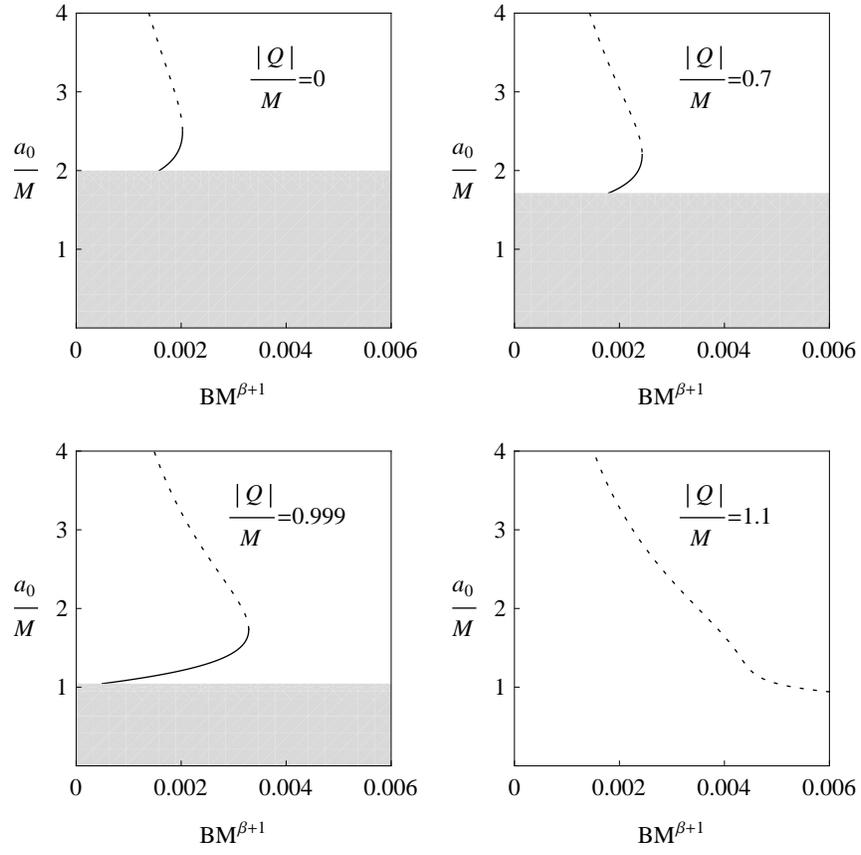}\caption{RN thin-shell wormholes for
$\beta=1,~A=1$ with different values of $\frac{|Q|}{M}$.}
\end{figure}

\subsection{Schwarzschild de-Sitter and anti de-Sitter Wormholes}

For the Schwarzschild de-Sitter and anti de-Sitter thin-shell
wormholes, the metric functions have the following form
\begin{eqnarray}\label{31}
N(r)=1-\frac{2M}{r}-\frac{\Lambda{r^2}}{3},\quad G(r)=r^2,
\end{eqnarray}
where $\Lambda>0$ is the cosmological constant. The function $N(r)$
remains negative for $\Lambda{M^2}>\frac{1}{9}$, while for
$0<\Lambda{M^2}\leq{\frac{1}{9}}$, the Schwarzschild de-Sitter
geometry has two horizons event $r_h$ and cosmological $r_c$ as
follows
\begin{eqnarray}\label{32}
r_h&=&\frac{-1+\dot{\iota}\sqrt{3}-(1+\dot{\iota}\sqrt{3})(-3\sqrt{\Lambda}M
+\dot{\iota}\sqrt{1-9{\Lambda}M^2})^\frac{2}{3}}{2{\sqrt{\Lambda}}(-3\sqrt{\Lambda}M
+\dot{\iota}\sqrt{1-9{\Lambda}M^2})^\frac{1}{3}}, \\\label{33}
r_c&=&\frac{1+(-3\sqrt{\Lambda}M
+\dot{\iota}\sqrt{1-9{\Lambda}M^2})^\frac{2}{3}}{{\sqrt{\Lambda}}(-3\sqrt{\Lambda}M
+\dot{\iota}\sqrt{1-9{\Lambda}M^2})^\frac{1}{3}}.
\end{eqnarray}
The horizons $r_h$ and $r_c$ are increasing and decreasing function
of $\Lambda$, since $\lim_{\Lambda\rightarrow{0^+}}r_h=2M$.
Moreover, $\lim_{\Lambda\rightarrow{0^+}}r_c=+\infty$ and
$r_h=r_c=3M$ for $\Lambda{M^2}=\frac{1}{9}$, which is not possible
in wormhole configuration. Thus the static solution will exist if
$0<\Lambda{M^2}<\frac{1}{9}$ and $r_h<a_0<r_c$. The event horizon
$r'_h$ for the Schwarzschild anti de-Sitter geometry is given by
\begin{eqnarray}\label{34}
r'_h=\frac{1-(-3\sqrt{|\Lambda|}M
+\sqrt{1+9{|\Lambda|}M^2})^\frac{2}{3}}{{\sqrt{|\Lambda|}}(-3\sqrt{|\Lambda|}M
+\sqrt{1+9{|\Lambda|}M^2})^\frac{1}{3}},
\end{eqnarray}
which is continuous and increasing function of $\Lambda$, as
$\lim_{\Lambda\rightarrow{0^-}}r'_h=2M$ and
$\lim_{\Lambda\rightarrow{-\infty}}r'_h=0$. This shows that $r'_h$
has values in the range $0<r'_h<2M$ and the static solution will
exist whenever $a_0>r'_h$. The corresponding surface energy density
and pressure at the throat become
\begin{eqnarray}\label{35}
\sigma_0=-\frac{\sqrt{3a_0-6M-\Lambda{a^3_0}}}{2\pi{a_0}\sqrt{3a_0}},\quad
p_0=\frac{3a_0-3M-2\Lambda{a^3_0}}{4\pi{a_0}\sqrt{3a_0(3a_0-6M-\Lambda{a^3_0})}}.
\end{eqnarray}
\begin{figure}
\centering \epsfig{file=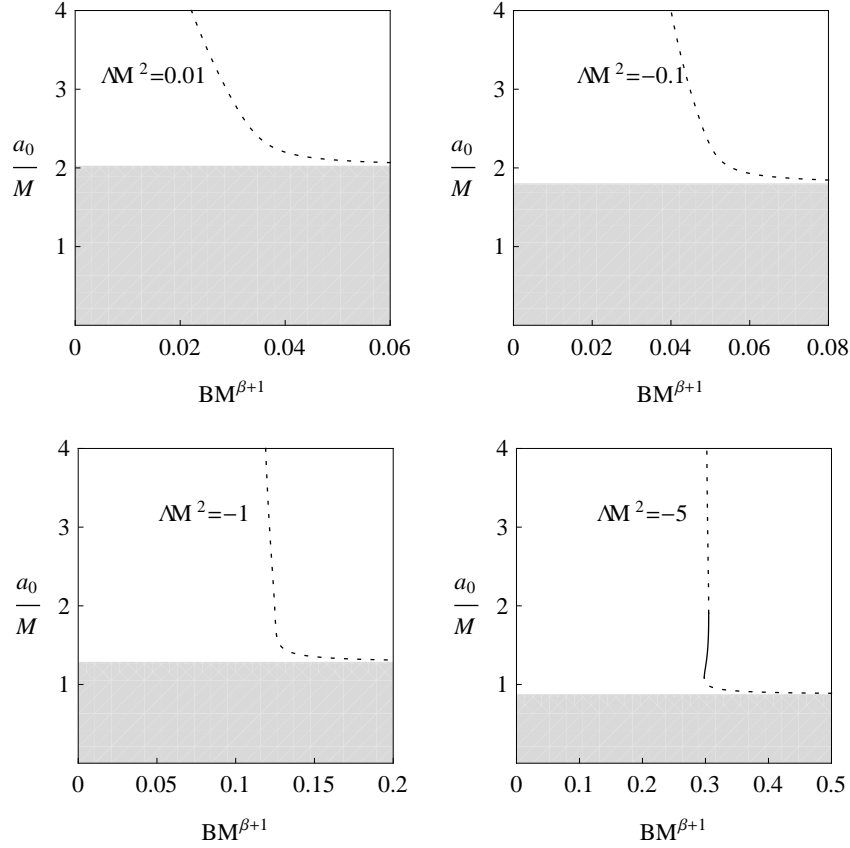}\caption{Schwarzschild de-Sitter and
anti de-Sitter thin-shell wormholes under radial perturbations with
parameters $\beta=0.2,~A=1,~\Lambda{M^2}$. The solid and dotted
curves indicate the stable and unstable solutions respectively, and
the grey zones correspond to the non-physical regions, where
$a_0\leq{r_h}$.}
\end{figure}
\begin{figure}
\centering \epsfig{file=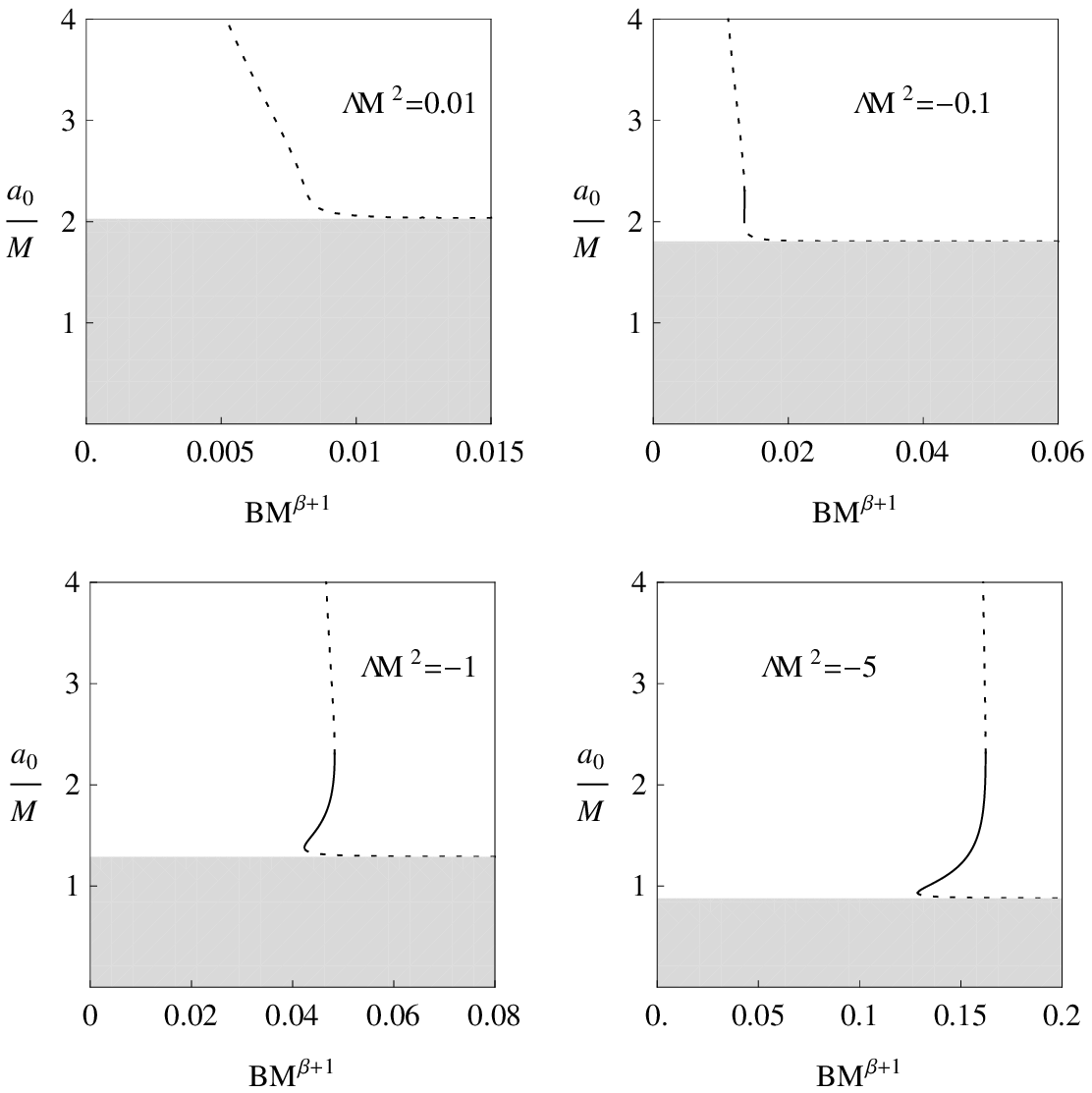}\caption{Schwarzschild de-Sitter and
anti de-Sitter thin-shell wormholes with
$\beta=0.6,~A=1,~\Lambda{M^2}$.}
\end{figure}
\begin{figure}
\centering \epsfig{file=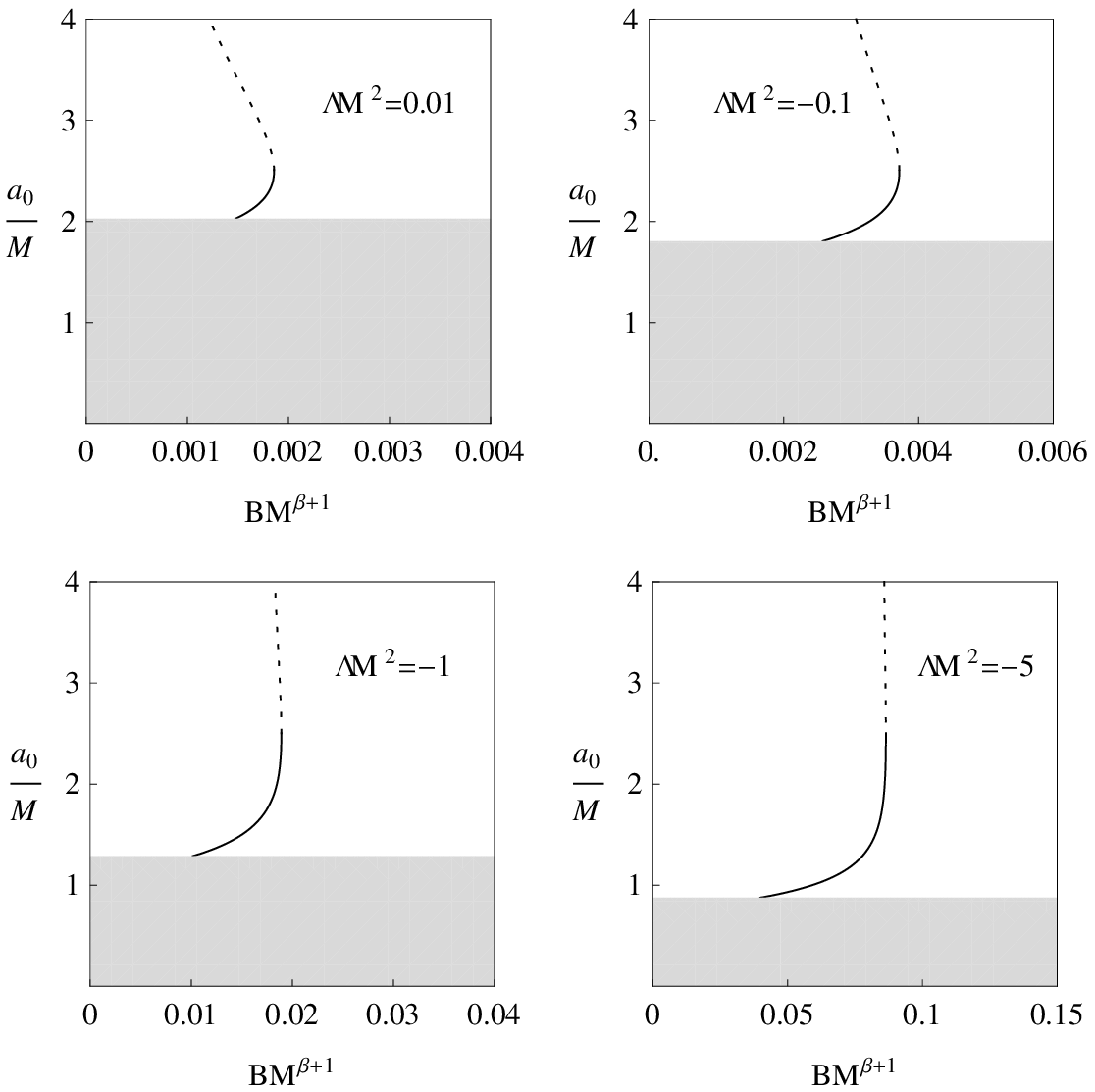}\caption{Schwarzschild de-Sitter and
anti de-Sitter thin-shell wormholes with
$\beta=1,~A=1,~\Lambda{M^2}$.}
\end{figure}

The dynamical equation and potential function from Eqs.(\ref{13})
and (\ref{25}) with (\ref{31}) yield
\begin{eqnarray}\nonumber
&&a_0(1+2A)-(1+4A)M-\frac{2}{3}\Lambda{a^3_0}(1+A)-2B(2\pi)^{1+\beta}a^{\frac{3(1+\beta)}{2}}_0
\\\label{36}&\times&(a_0-2M-\frac{\Lambda}{3}a^3_0)^{\frac{1-\beta}{2}}=0.
\end{eqnarray}
\begin{eqnarray}\nonumber
\Phi''(a_0)&=&\frac{2}{a^3_0(\Lambda{a^3_0}-3a_0+6M)}\left
\{(1+\beta)\left[3a^2_0(1+2A)+9M^2(1+4A)\right.\right.
\\\nonumber&-&\left.\left.30a_0AM+6a^3_0{\Lambda}AM-2A{\Lambda}a^4_0\right]
-3a_0M(3+4\beta)\right.\\\label{37}&-&\left.2a^4_0{\beta}\Lambda
+3a^3_0{\Lambda}M(2\beta-1)\right\}.
\end{eqnarray}
Similar to the RN case, we solve Eq.(\ref{36}) numerically for $a_0$
and explore its stability. The results are shown in Figures
\textbf{4-6} for $0<\beta\leq{1}$ with different values of
parameters.
\begin{itemize}
\item{In Figure \textbf{4} when $\beta=0.2$, there exists one unstable static solution
corresponding to $\Lambda{M^2}=0.01,~-0.1,~-1$, while two unstable
and one stable solution for $\Lambda{M^2}=-5$. Also, the throat
radius decreases and touches the horizon radius for large value of
$BM^{\beta+1}$.}
\item{When $\beta=0.6$ (Figure \textbf{5}), one unstable static solution corresponds to $\Lambda{M^2}=0.01$,
one stable and two unstable for $\Lambda{M^2}=-0.1,~-1,~-5$.}
\item{In Figure \textbf{6} $(\beta=1)$, there exist
both stable and unstable solutions for each case corresponding to
$\Lambda{M^2}=0.01,-0.1,~-1,-5$.}
\end{itemize}

In this case, we have also found some extra solutions for the MCG.
Notice that all the solutions are of unstable type for $\beta=0.2$
with different values of $\Lambda{M^2}$ \cite{15}. On the other
hand, we have found two extra stable static solutions for
$\Lambda{M^2}=-5$. Similarly, for $\beta=0.6$, there exist two extra
stable and unstable solutions for $\Lambda{M^2}=-0.1,~-1$. For
$\beta=1$, we have two extra stable solutions for
$\Lambda{M^2}=0.01,~-0.1$, whereas for $\Lambda{M^2}=-1,~-5$, we
have similar solutions to \cite{15}.

\section{Conclusions}

In this paper, we have found a class of spherical thin-shell
wormholes with cut and paste technique. We have assumed MCG to deal
with the exotic matter located in the wormhole throat $\Sigma$. The
Darmois-Israel formalism has been used to find the surface energy
density and pressure. We have manipulated the results numerically
and adopted the standard potential approach to investigate stability
of the static wormhole solutions. The results representing the
stable and unstable solutions with solid and dotted curves,
respectively are shown in Figures \textbf{1-6}. In particular, we
have explored stability of RN as well as Schwarzschild de-Sitter and
anti-de-Sitter thin-shell wormholes for different values of the gas
exponent $\beta$ and compared our results with those already found
in \cite{15} for the generalized Chaplygin gas. It is concluded that
some extra stable as well as unstable static solutions exist
depending upon the parameters
$A,~B,~\beta,~\frac{|Q|}{M},~\Lambda{M^2}$ involving in the model.
This shows that the choice of equation of state plays a vital role
in the existence of wormhole solutions. We would like to mention
here that all our results reduce to that of \cite{15} when we take
$A=0$.

\vspace{0.5cm}

{\bf Acknowledgments}

\vspace{0.5cm}

We would like to thank the Higher Education Commission, Islamabad,
Pakistan, for its financial support through the {\it Indigenous
Ph.D. 5000 Fellowship Program Batch-VII}. One of us (MA) would like
to thank University of Education, Lahore for the study leave.

\end{document}